\documentclass[aps,pra,superscriptaddress,a4paper,reprint]{revtex4-1}

\usepackage[utf8]{inputenc}
\usepackage{amsmath}
\usepackage{graphicx}
\usepackage[left=2cm,right=2cm,
    top=2.5cm,bottom=2.5cm,bindingoffset=0cm]{geometry}
\usepackage{ulem}

\usepackage{hyperref}
\hypersetup
{%
pdftitle = {Focusing characteristics of a 4pi parabolic mirror light-matter interface},
colorlinks = {false}
}

\begin{document}

\title{Focusing characteristics of a 4$\pi$ parabolic mirror light-matter interface}

\author{Lucas Alber}
\thanks{These authors contributed equally to this work.}
\affiliation{Max Planck Institute for the Science of Light,
  Guenther-Scharowsky-Str. 1/ building 24, 91058 Erlangen, Germany}
\email{lucas.alber@mpl.mpg.de}
\affiliation{Friedrich-Alexander-Universit\"at Erlangen-N\"urnberg (FAU),
  Department of Physics, Staudtstr. 7/B2, 91058 Erlangen, Germany}

\author{Martin Fischer}
\thanks{These authors contributed equally to this work.}
\affiliation{Max Planck Institute for the Science of Light,
  Guenther-Scharowsky-Str. 1/ building 24, 91058 Erlangen, Germany}
\affiliation{Friedrich-Alexander-Universit\"at Erlangen-N\"urnberg (FAU),
  Department of Physics, Staudtstr. 7/B2, 91058 Erlangen, Germany}

\author{Marianne Bader}
\affiliation{Max Planck Institute for the Science of Light,
  Guenther-Scharowsky-Str. 1/ building 24, 91058 Erlangen, Germany}
\affiliation{Friedrich-Alexander-Universit\"at Erlangen-N\"urnberg (FAU),
  Department of Physics, Staudtstr. 7/B2, 91058 Erlangen, Germany}

\author{Klaus Mantel}
\affiliation{Max Planck Institute for the Science of Light,
  Guenther-Scharowsky-Str. 1/ building 24, 91058 Erlangen, Germany}

\author{Markus Sondermann}
\affiliation{Max Planck Institute for the Science of Light,
  Guenther-Scharowsky-Str. 1/ building 24, 91058 Erlangen, Germany}
\affiliation{Friedrich-Alexander-Universit\"at Erlangen-N\"urnberg (FAU),
  Department of Physics, Staudtstr. 7/B2, 91058 Erlangen, Germany}

\author{Gerd Leuchs}
\affiliation{Max Planck Institute for the Science of Light,
  Guenther-Scharowsky-Str. 1/ building 24, 91058 Erlangen, Germany}
\affiliation{Friedrich-Alexander-Universit\"at Erlangen-N\"urnberg (FAU),
  Department of Physics, Staudtstr. 7/B2, 91058 Erlangen, Germany}
\affiliation{Department of Physics, University of Ottawa, Ottawa,
  Ont. K1N 6N5, Canada}


\begin{abstract} 
Focusing with a 4$\pi$\ parabolic mirror allows for concentrating
light from nearly the complete solid angle, whereas focusing with a
single microscope objective limits the angle cone used for focusing to
half solid angle at maximum. Increasing the solid angle by using deep
parabolic mirrors comes at the cost of adding more complexity to the
mirror's fabrication process and might introduce errors that reduce
the focusing quality. To determine these errors, we experimentally
examine the focusing properties of a 
4$\pi$\ parabolic mirror that was produced by single-point diamond 
turning. The properties are characterized with a single 
$^{174}$Yb$^{+}$ ion as a mobile point scatterer. The ion is
trapped in a vacuum environment with a movable high optical access Paul
trap. We demonstrate an effective focal spot size of 209\,nm in lateral
and 551\,nm in axial direction. Such tight focusing allows us to build
an efficient light-matter interface. Our findings agree with numerical
simulations incorporating a finite ion temperature and
interferometrically measured wavefront aberrations induced by the
  parabolic mirror. We point at further technological improvements and
  discuss the general scope of applications of a 4$\pi$\ parabolic mirror.
\end{abstract}

\maketitle

\section{Introduction}
Free space interaction between light and matter is incorporated as a
key technology in many fields in modern science. The efficiency of
interaction influences measurements and applications ranging from
various kinds of fundamental research to industrial applications. New
innovations and new types of high precision measurements can be
triggered by improving the tools needed for a light-matter
interface. To achieve high interaction probability with a focused
light field in free space, an experimental scheme using parabolic
mirrors for focusing onto single atoms has been developed in recent
years
\cite{Quabis_OptComm2000,sondermann_design_2007,0295-5075-86-1-14007}. This
scheme relies on mode matching of the focused radiation to an 
electric dipole mode (cf. Ref.\,\cite{L_M_interaction} and citations therein). 

Focusing in free-space experiments is usually done with
state-of-the-art lens based imaging systems\,\cite{piro_heralded_2011,
  Kurtsiefer_Lens, Sandoghar_Nat_Photons,
  Imamoglu_single_photons}. Single lenses, however, suffer from
inherent drawbacks like dispersion induced chromatic aberrations,
optical aberrations, and auto-fluorescence, respectively. Most of
these limitations can be corrected to a high degree by precisely
assembling several coated lenses in a lens-system, \textit{e.g.} in a 
high numerical aperture (\textit{NA}) objective. Although solving some
problems, multi-lens-systems induce new problems such as short working
distances, low transmission for parts of the optical spectrum, the
need for immersion fluids, and high costs, respectively.
Therefore, multi-lens systems are often application specific
providing best performance only for the demands that are most
important for the application. 

Mirror based objectives are an alternative to lens-based systems and
can overcome some of these problems. The improvement is based on a
mirror's inherent property of being free from chromatic
aberrations. The nearly wavelength independent behavior also leads to a
homogeneous reflectivity for a large spectral window. Comparing the
reflectivity of mirrors to the transmission of lens based objectives,
mirrors can sometimes also surpass lens-based systems.
But surprisingly, they are rarely used when high interaction
efficiency is required.
This lack in application may be due to the fact that
reflecting imaging systems, like the Cassegrain reflector, cannot
provide a high \textit{NA}. A high \textit{NA} is however needed for
matching the emission pattern of a dipole, which spans over the entire
solid angle. The limitation in \textit{NA} consequently constitutes a
limitation in the maximum achievable light-matter coupling
efficiency. 

High NA parabolic mirrors (\textit{NA} = 0.999) have meanwhile been
successfully applied as objectives in confocal
microscopy\,\cite{drechsler2001,stadler2008}, demonstrating the
potential for imaging applications. The parabolic mirror (PM) is a
single optical element that, in theory, can cover nearly the complete
4$\pi$ solid angle for tight focusing\,\cite{lindlein_new_2007}. In
this article we report on the detailed characterization of such
a 4$\pi$ parabolic mirror (4$\pi$-PM), in which we sample the focal
intensity distribution with a single $^{174}$Yb$^+$ ion, 
trapped in a stylus like movable Paul trap\,\cite{maiwald_stylus_2009}.

In contrast to our previous studies\,\cite{fischer_efficient_2014},
we measure the response of the ion at a wavelength
different to the one used for excitation.
This approach is standard in fluorescence microscopy and has
also been used in experiments with trapped ions\,\cite{linke2012}.
It renders unnecessary a spatial separation of focused light and light
scattered by the ion, thus lifting the limitation of
focusing only from half solid angle as in
Ref.\,\cite{fischer_efficient_2014}.
However, we will find below that by using the solid angle provided by
our 4$\pi$-PM the measured effective excitation point
spread function (\textit{PSF}) is worse when using the full mirror as
compared to focusing from only half solid angle.
As outlined below, this is not a general restriction but specific to
the aberrations of the mirror used in our experiments.
It is a challenge to determine the aberrations of such a deep
parabolic mirror\,\cite{leuchs_interferometric_2008} and we discovered
the full extent of these aberrations only when scanning the 3D field
distribution with the single ion, revealing an error in the earlier
interferometric measurements. 
Here, we present a reasonable agreement of the experiments with
results of simulations incorporating a finite ion temperature and new
interferometrically measured wavefront aberrations of the parabolic
mirror itself.

Despite of these aberrations, the efficiency obtained here for
coupling the focused light to the linear dipole transition of the
$^{174}$Yb$^+$ ion is better than reported
previously\,\cite{fischer_efficient_2014}, using the full mirror 
as well as focusing from half solid angle.
As a further improvement in comparison to
Ref.\,\cite{fischer_efficient_2014} we keep the excitation of the
ion well below saturation making sure that the size of the ion's wave function
stays approximately constant as much as possible. 
All in all, the ion constitutes a nanoscopic probe with
well defined properties throughout the measurement range.

In the concluding discussion of this paper,
the parabolic mirror is compared to
other high \textit{NA} focusing tools, especially to lens-based
4$\pi$\ microscopes. Its possible field of application is discussed
and further improvements to the existing set-up are proposed.

\section{Setup and experiment}

Our main experimental intention is to focus light to a minimal spot
size in all spatial directions simultaneously. The highest electric
energy density that can be realized with any focusing optics is
created by an electric dipole wave\,\cite{bassett_limit_1986}.
We therefore choose this type of spatial mode in our experiment. The
electric dipole wave is created by first converting a linear
polarized Gaussian beam into a radially polarized donut mode via a
segmented half-wave plate (B-
Halle)\,\cite{quabis_generation_2005,golla2012}.
Second, the radially polarized donut
mode is focused with a parabolic mirror onto the trapped ion. This, in
theory, enables us to convert approximately $91\,\%$ of the donut mode
into a linear dipole mode\,\cite{SondermannArXiv2008}. The conversion
efficiency is limited since the donut mode is only approximating the
ideal spatial mode that is necessary to create a purely linear dipole
mode\,\cite{sondermann_design_2007, PhysRevA_GAlber} by being focused
with the parabolic mirror. The donut mode, however, yields the
experimental advantage of being propagation invariant and comparably
  easy to generate. 

Our focusing tool, the parabolic mirror, is made of diamond turned
aluminum (Fraunhofer Institute for Applied Optics and Precision
Engineering, Jena) with a reflectivity of 64\,\% for the incident mode
at a wavelength of $\lambda_{exc} = 369.5$\,nm. Its geometry has a
focal length of 2.1\,mm and an outer aperture of 20\,mm in diameter.
In total, the geometry covers 81\,\% of the complete solid angle. This
fraction corresponds to 94\,\% of the solid angle that is relevant for
coupling to a linear dipole oriented along the axis of symmetry
\cite{SondermannArXiv2008,lindlein_new_2007}. Furthermore, the mirror
has three bores near its vertex: two bores with a diameter of 0.5\,mm
for dispensing neutral atoms and for illuminating the ion with
additional laser beams, respectively; and one bore with a diameter of
1.5\,mm for the ion trap itself. The ion trap is a Stylus-like Paul
trap similar to\,\cite{maiwald_stylus_2009} with high optical
access. The trap is mounted on a movable xyz piezo translation stage
(PIHera P-622K058, Physik Instrumente) that is used for measuring
the effective excitation \textit{PSF}. The effective excitation
\textit{PSF} is defined as the convolution of the focal intensity
distribution with the spatial extent of the ion.   

\begin{figure}
\includegraphics[width=0.95\linewidth]{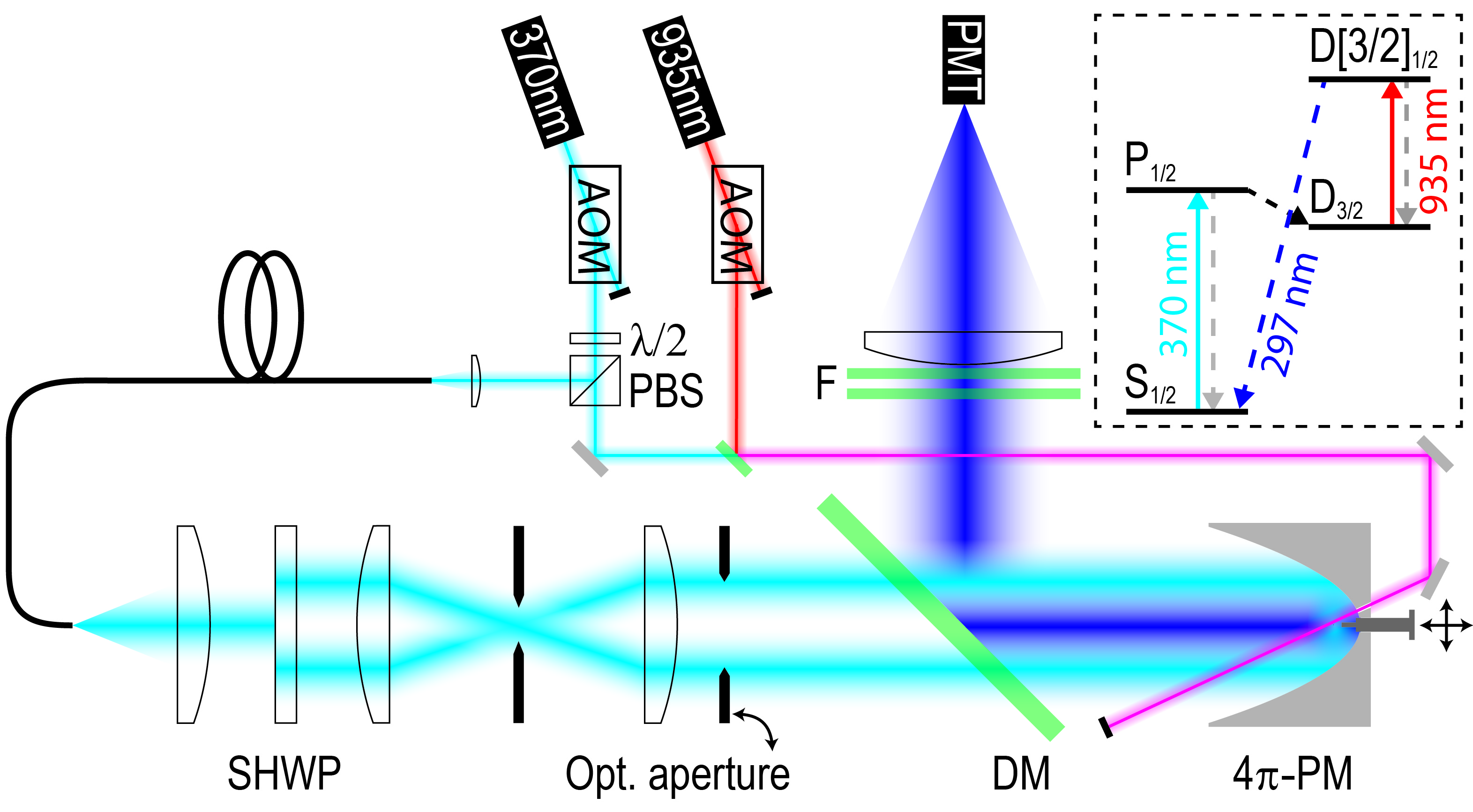}
\caption{\label{fig:setup} 
  Optical set-up of the experiment and relevant energy levels of
  $^{174}$Yb$^+$. The cooling laser (blue) and the repump laser
  (red) are focused onto the ion through a hole at the backside of the
  parabolic mirror. AOM - acousto optical modulator,
  DM - dichroic mirror, F - clean up filter,
  4$\pi$-PM - 4$\pi$ parabolic mirror, PMT - photo multiplier
  tube, SHWP - segmented half-wave plate.} 
\end{figure}

We measure the effective excitation \textit{PSF} by probing the focal
spot at different positions. In order to do so, we use the translation
stage to scan the ion through the focal spot with an increment of
25\,nm. At each position, the incoming dipole mode excites the
$\text{S}_{1/2}\text{ - P}_{1/2}$ transition that has a wavelength of
$\lambda_{exc} = 369.5$\,nm and a transition linewidth of $\Gamma/2\pi
= 19.6$\,MHz\,\cite{meyer_laser_2012}. The relevant energy levels of 
$^{174}\text{Yb}^+$ are shown in figure \ref{fig:setup}. We weakly
drive this transition such that the probability for exciting the ion
into the $\text{P}_{1/2}$ state is proportional to the electric energy
density at any point in the focal area. During these measurements, the
ion is Doppler cooled by the 
  focused donut mode. Hence, the detuning of this mode relative to the
  $\text{S}_{1/2}\text{ - P}_{1/2}$ transition and its power determine
  the temperature of the ion, see appendix for further details.

\begin{table}[h!]
\centering
\begin{tabular}{lcc}
\hline\hline
Transition&Branching ratio&Decay rate [$\Gamma/2\pi$]\\
\hline
$^2 \text{P}_{1/2}\text{ - } ^2 \text{S}_{1/2}$&$99.5\,\%$&$19.6$ MHz\\
$^2 \text{P}_{1/2}\text{ - } ^2 \text{D}_{3/2}$&$0.5\,\%$& \\
$^3 \text{D[3/2]}_{1/2}\text{ - } ^2 \text{D}_{3/2}$&$1.8\,\%$&$4.2$ MHz\\
$^3 \text{D[3/2]}_{1/2}\text{ - } ^2 \text{S}_{1/2}$&$98.2\,\%$& \\
$^2 \text{D}_{3/2}\text{ - } ^2 \text{S}_{1/2}$& &$3$ Hz\\
\hline\hline
\end{tabular}
 \caption{  \label{tab:Transitions}
Branching ratios and decay rates for the relevant transitions of
$^{174} \text{Yb}^{+}$ taken from\,\cite{meyer_laser_2012,
  olmschenk_manipulation_2007} and citations therein.}  
\end{table}
  
Since we excite the ion from the complete solid angle that is covered
by the 4$\pi$\ parabolic mirror and since almost all excitation light
is reflected into the detection beam path by the parabolic mirror, we
cannot directly detect the 
fluorescent response of the ion at the same
wavelength. Instead, we detect photons at a wavelength of
$\lambda_{det} = 297.1$\,nm allowing us to independently focus and
detect from nearly the complete solid angle. Photons at the detection
wavelength $\lambda_{det}$ are emitted during the spontaneous
$\text{D[3/2]}_{1/2}\text{ - S}_{1/2}$ decay
\cite{meyer_laser_2012}. The $\text{D[3/2]}_{1/2}$ level is populated
when the ion spontaneously decays from the excited $\text{P}_{1/2}$
state into the metastable $\text{D}_{3/2}$ state (branching ratio $\beta = 0.5\,\%$,
lifetime of 52\,ms\,\cite{olmschenk_manipulation_2007}, see table
\ref{tab:Transitions}). From this 
state, we optically pump the ion into the $\text{D[3/2]}_{1/2}$ state
by saturating the $\text{D}_{3/2}\text{ - D[3/2]}_{1/2}$ transition
with a strong laser field at a wavelength of 935.2\,nm (DL-100,
Toptica Photonics). The upper state of this transition decays to the
$\text{S}_{1/2}$ ground state with a probability of 98\,\%
\cite{meyer_laser_2012}. The infrared laser is co-aligned with a
second 
laser at the excitation wavelength $\lambda_{exc}$ (TA-SHG pro,
Toptica Photonics) and both are sent through the focus of the
parabolic mirror via one of its backside bores (see figure
\ref{fig:setup}). The second laser at the excitation wavelength
$\lambda_{exc}$ is used for ionization. 

The emitted fluorescent photons at the detection wavelength
$\lambda_{det}$ are out-coupled from the excitation beam path via a
dichroic mirror (FF310-Di01, Semrock) and two clean up filters
(FF01-292/27-25, Semrock). Afterwards, we detect them with a
photomultiplier tube in Geiger mode operation (MP-942, Perkin Elmer)
that has a remaining underground/dark count rate of 10 - 20\,cps. The
overall detection efficiency $\eta_{det}$ at the detection wavelength
$\lambda_{det}$ was measured via pulsed excitation and amounts to
$\eta_{det} \approx 1.4\,\%$ (see appendix). The detection efficiency is needed for
the determination of the coupling efficiency to the trapped
ion.

Based on the atomic decay rate on the detected transition, the total
photon emission rate would be approximately R =
$\beta\,\frac{\Gamma}{2}\frac{1}{2} = 154\,\text{kcps}$ for $S = 1$
(see equation \ref{eq:DetRate}). Taking into account the finite
detection efficiency, we would expect to measure approximately
$R_{det} = \eta_{det}\,154\,\text{kcps} = 2160\,\text{cps}$.

The coupling efficiency is measured by recording the detection count
rate $R_{det}$ as a function of the excitation power
$P_{exc}$. Analyzing the four-level quantum master equation we find
that both quantities are proportional to each other in the limit of
strong repumping powers and saturation parameters $S\ll1$. The
latter condition is met by keeping $S\le0.1$ in our measurements.
This also ensures that the spatial extent of the ion is approximately
constant throughout the measurement, see appendix.
The dependence of $R_{det}$ for varying excitation power is given by
\begin{align}
  R_{det}& = \eta_{det}\,\beta\,\frac{\Gamma}{2}\frac{S}{S+1} \nonumber\\
  & = \eta_{det}\,\beta\,\frac{\Gamma}{2}
  \frac{G{P_{exc}}/{P_{sat}}}{G{P_{exc}}/{P_{sat}}+1} 
	\label{eq:DetRate}
\end{align}
with $S$ denoting the saturation parameter, and $G$ the coupling
efficiency, respectively. The saturation power $P_{sat}$ is defined as
$P_{sat}=3\cdot\frac{h c} 
  {\lambda_{exc}}\frac{\Gamma}{8}(1+4({\Delta}/{\Gamma})^2)$.
The factor $3$ accounts for the fact that we are
not driving a closed linear-dipole transition but a J=1/2
$\leftrightarrow$ J=1/2 one. 
The relation between saturation parameter, saturation power and
  coupling efficiency is $S=G{P_{exc}}/{P_{sat}}$\,\cite{fischer_efficient_2014}. 
$\Delta$ is the detuning of the excitation laser from
the $\text{S}_{1/2}\text{ - P}_{1/2}$ resonance. The formula for the
detection count rate enables us to determine the coupling efficiency
by curve fitting of our measured data for $R_{det}$ as a function of
$P_{exc}$. For the curve fitting, all parameters except the coupling
efficiency are kept constant. During the measurement of the 
coupling efficiency, we position the ion exactly in the maximum of the
excitation \textit{PSF}, \textit{i.e.} we measure the maximum coupling
efficiency obtainable in the focal region under the current experimental conditions.

\section{Results}

\begin{figure}
\centering
\includegraphics[width=0.95\linewidth]{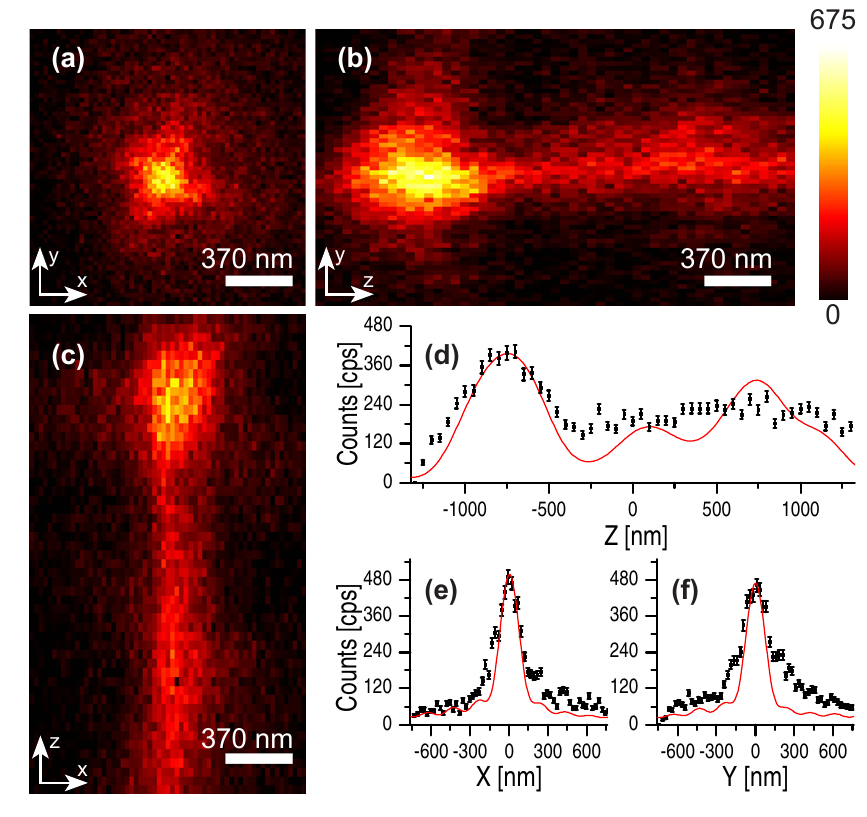}
\caption{\label{fig:PSF_fSA} 
  Effective excitation \textit{PSF} in the xy (a), zy (b), and
  xz (c) plane when illuminating the \textit{full solid angle} covered
  by the 4$\pi$-PM. The corresponding line profiles through the center
  of the focus are shown in (d-f).  They are overlayed to the line
  profiles resulting from numerical simulations (red)} 
\end{figure}

\begin{figure}
\centering
\includegraphics[width=0.95\linewidth]{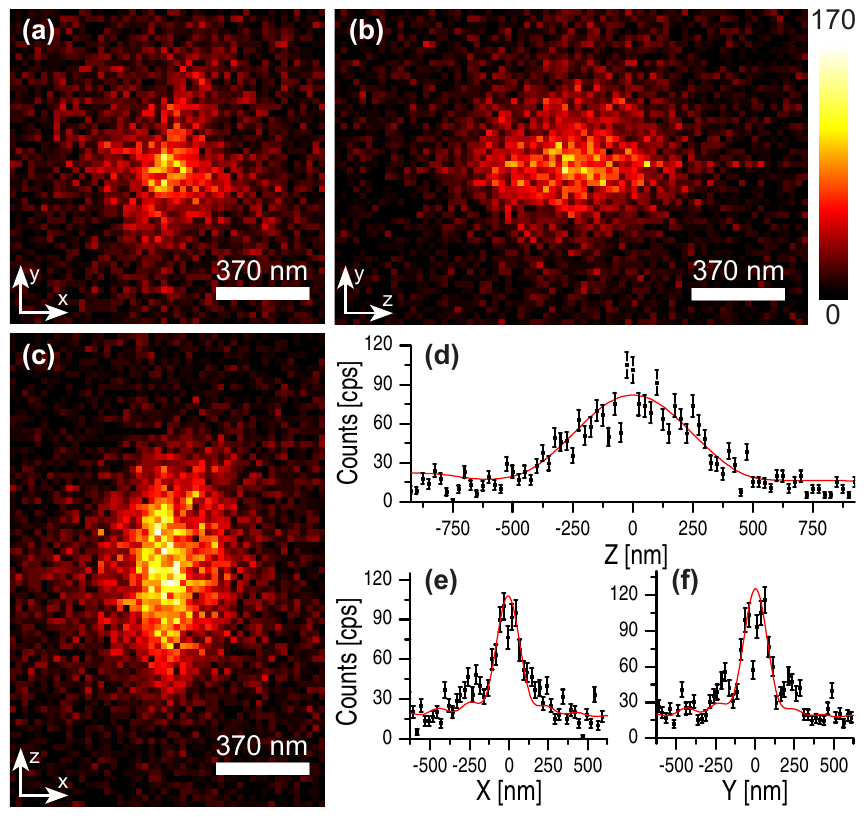}
\caption{\label{fig:PSF_hSA}
  Effective excitation \textit{PSF} similar to figure
  \ref{fig:PSF_fSA} but for illuminating \textit{half of the solid
    angle} of the 4$\pi$-PM. This is accomplished by using the
  optional aperture shown in figure \ref{fig:setup}.} 
\end{figure}

The experimental results for the effective excitation \textit{PSF} are
shown in figure \ref{fig:PSF_fSA}. We measure a spot size of $237\pm 10$\,nm
(\textit{FWHM}) in the lateral direction (a, e, f). In the axial
direction (b - d), however, the focal peak is broadened due to optical
aberrations. The influence of the aberrations is reduced, when
we limit the front aperture of the 4$\pi$-PM to half solid angle
(figure \ref{fig:PSF_hSA}). The reduced aperture results in a lateral
width of $209\pm 20$\,nm and an axial width of $551\pm 27$\,nm. These values include
the influence of the finite spatial extent of the trapped ion (see
appendix). In the Doppler limit, this extent is approximately 140\,nm
in lateral and 80\,nm in axial direction considering the trap
frequencies $\omega_{\text{lateral}} /2\pi \cong 490$\,kHz and
$\omega_{\text{axial}} /2\pi\cong 1025$\,kHz, respectively, and a
detuning from   resonance of about 14.1 MHz.

To determine the minimal contribution of the ion-size to the focal
broadening when Doppler cooling, we 
simulate the excitation \textit{PSF} based on a generalization of the
method presented in\,\cite{richards_electromagnetic_1959}. Our
simulation also includes the aberrations of the parabolic mirror which
were measured interferometrically
beforehand\,\cite{leuchs_interferometric_2008}.
The intensity distributions resulting from simulations only accounting
for mirror aberrations are subsequently convolved with the spatial
extent of the ion to achieve the effective excitation \textit{PSF}. 

\begin{figure}
\centering
\includegraphics[width=0.95\linewidth]{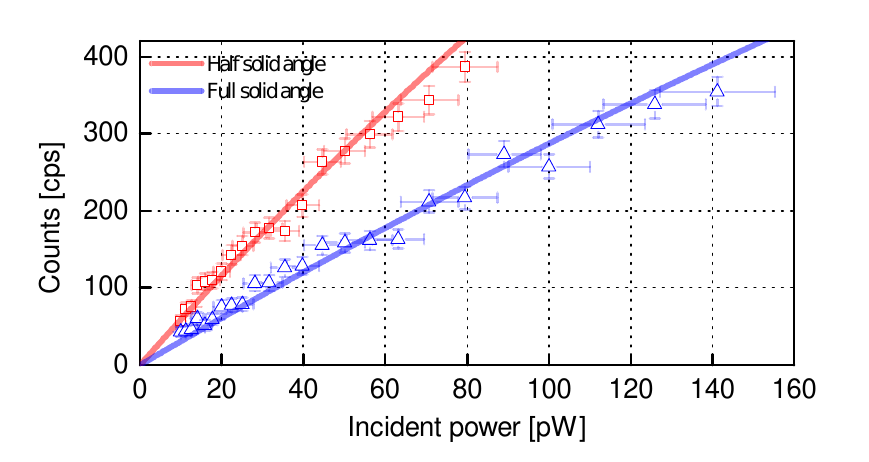}
\caption{\label{fig:Saettigung}
  Measurement of the ion's fluorescence vs. excitation power for
  determining the coupling efficiency (symbols).
  Solid lines show the results from curve fitting when illuminating
  half of the solid angle (red) and full solid angle (blue). To ensure
  constant coupling for the curve fitting, we only use data points for
  weak excitation ($R_{det}\leq 392$\,cps equals $S < 1/10$, see
  appendix) which corresponds to the nearly linear region of
  a saturation curve.} 
\end{figure}

\begin{table*}
\centering
\begin{tabular}{llcccc}
\hline\hline
\multicolumn{2}{c}{Simulation} & \multicolumn{2}{c}{Lateral} & \multicolumn{2}{c}{Axial}\\
 & & HSA & FSA &HSA&FSA\\
\hline
Exc. \textit{PSF} & ideal mirror (i.m.)& 139 nm & 142 nm & 412 nm & 253 nm\\
& aberrated mirror (a.m.) & 135 nm & 135 nm & 536 nm & \\
& i.m. with ion-extent& 189 nm & 192 nm & 418 nm & 266 nm\\
& a.m. with ion-extent & 189 nm & 189 nm & 542 nm & \\
\hline \hline
\multicolumn{2}{l}{Eff. exc. \textit{PSF} (measurement)} & $209\pm 20$ nm & $237\pm 10$ nm & $551\pm 27$ nm & \\
\hline\hline
\end{tabular}
  \caption{  \label{tab:IonSize}
Influence of experimental factors on the effective
    excitation \textit{PSF} (eff. exc. \textit{PSF}). The full width at half maximum is given
    for illuminating full solid angle (FSA) and half solid angle (HSA)
    of the 4$\pi$-PM, respectively. In case of illuminating the full
    solid angle and considering optical aberrations, no distinct peak
    can be identified along the axial direction.} 
\end{table*}

The outcomes of our simulations are shown in
table\,\ref{tab:IonSize}. The effective \textit{PSF} obtained in the
simulations exhibits a good qualitative agreement with the
\textit{PSF} obtained in the experiment, cf. figure \ref{fig:PSF_fSA}
and \ref{fig:PSF_hSA}. Based on these results, the coupling 
efficiency $G$ is expected (see appendix) to be $G=8.7\,\%$ for
illuminating the full aperture and $G=14.3\,\%$ for limiting the
aperture of the 4$\pi$-PM to half solid angle. From the data shown in figure
\ref{fig:Saettigung}, we measure a coupling efficiency of $G=8.6\pm
0.9\,\%$ (full solid angle) and $G=13.7\pm 1.4\,\%$ (half solid
angle). These values are close to the expected values from the
simulations taking into account all current deficiencies of the set-up.

\section{Discussion}

Concentration of light into a narrow three dimensional volume is
involved in many scientific applications. The scope ranges from
applications that require "classical" light fields, like light
microscopy, optical traps and material processing, to applications in
quantum information science. 
In quantum information science, tight
focusing of light is the key ingredient for free-space light-matter
interfaces with a high coupling efficiency. This kind of free-space
set-up may be an alternative to cavity based light-matter interfaces
also providing high interaction strength. But in contrast to cavity
assisted set-ups, free-space experiments often have a low level of
instrument complexity and provide higher bandwidth. This is important
considering the scalability and flexibility of an experimental set-up. 

Technically, concentration of light is done by using focusing
optics. How tight the focusing will be, depends on the numerical
aperture of the focusing optics that is given by 
${\textit{NA}\,=\,n\,\sin(\alpha),\,\textit{NA}\in[0,n]}$, where
$\alpha$ is the half-opening angle of the optics' aperture and $n$ is
the refractive index of the surrounding medium. For high \textit{NA}
objectives, the exact dependence of the focal volume as a function of
\textit{NA} can only 
be calculated numerically incorporating a vectorial treatment of the
electric field. But as the numerical aperture increases, the focal
volume will basically decrease.  

\begin{figure}
\centering
\includegraphics[width=0.95\linewidth]{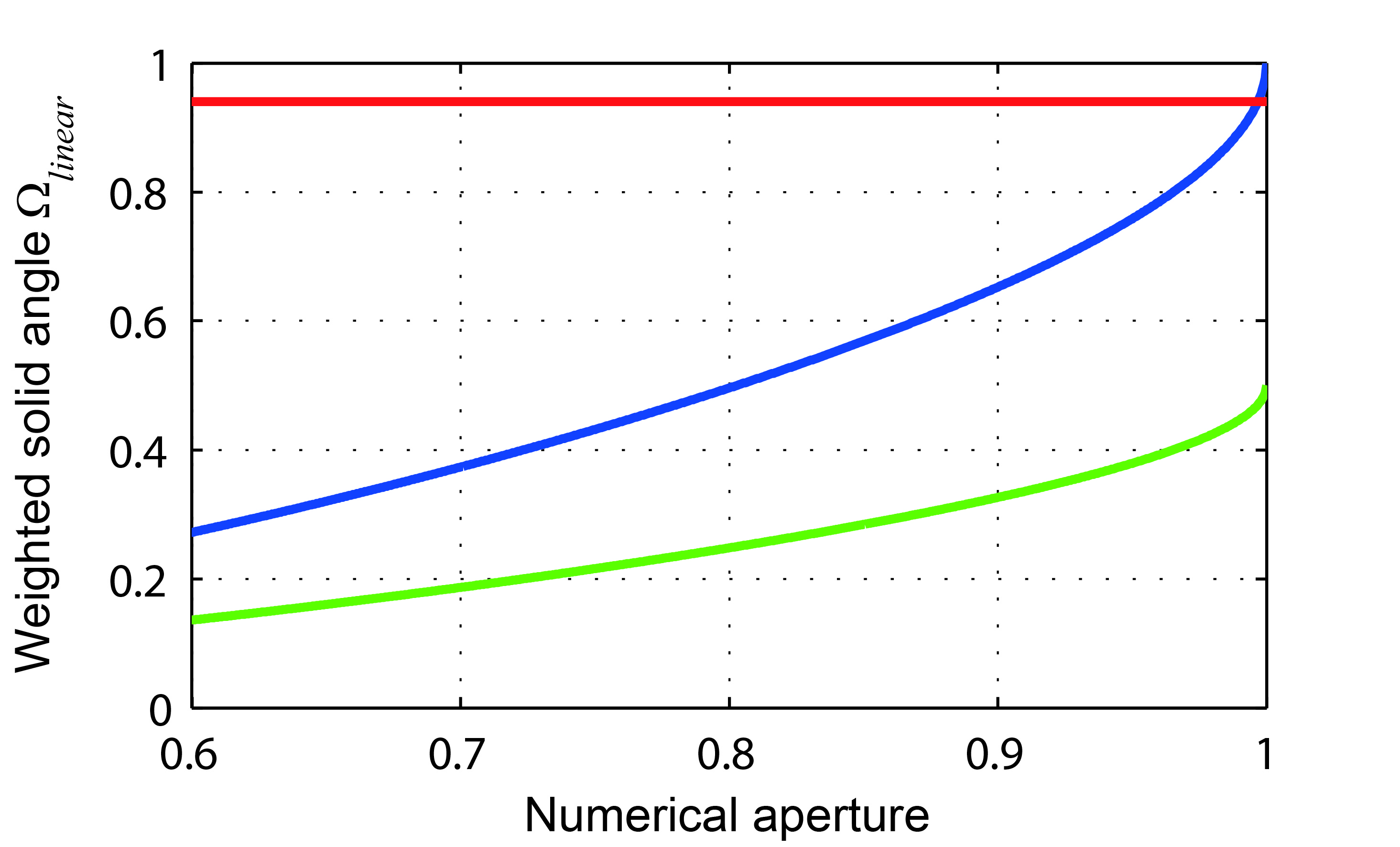}
\caption{\label{fig:epsilonNA}
  Weighted solid angle $\Omega_{linear}$ covered by a single
  objective lens (green), by a lens based 4$\pi$-microscope (blue),
  and by the 4$\pi$-PM geometry used in the experiment (red).} 
\end{figure}

For a 4$\pi$\ focusing optic, the numerical aperture is no longer
defined. Nevertheless, it seems obvious that diffraction limited
focusing from more than half of the hemisphere would produce a smaller
three-dimensional focal volume. Consequently, a different quantity has 
to be used for comparing the performance of different 4$\pi$\ focusing
optics. One suitable quantity is the weighted solid angle $\Omega \in
[0,1]$ (normalized to $8\pi/3$)\,\cite{SondermannArXiv2008}. It is the
solid angle that is covered by the focusing optics weighted by the
dipole's angular irradiance 
pattern. $\Omega$ defines the maximum fraction of incident power that
can be coupled into the dipole mode of a single emitter. Consequently,
it provides information 
about the ability to concentrate light since an electric dipole mode
achieves the highest possible energy concentration
\cite{bassett_limit_1986}. $\Omega = 1$ therefore means, that all of
the light is coupled into the dipole mode, assuming the ideal
radiation pattern. The focusing capabilities of such a focusing optics
can not be exceeded by any 
other optics. In figure \ref{fig:epsilonNA}, the maximal conversion
efficiency into a linear dipole wave $\Omega_{linear}$ is compared for
different (4$\pi$) focusing systems also including the 4$\pi$-PM
geometry used in the experiment. In case of our 4$\pi$-PM, one has
$\Omega_{linear} = 0.94$. The same fraction of the weighted solid
angle can be covered using two opposing 
objective lenses each having a \textit{NA} of 0.997 in vacuum. High
quality objectives of such high numerical aperture are, however, not
available.   

In the experiment, the weighted solid angle covered by the focusing
optics is one quantity that determines the measured light-matter
coupling efficiency. Other important experimental factors are the
optical aberrations and the spatial extent of the ion,
respectively. Under ideal conditions, we expect a coupling efficiency
of $G = \eta^2 \cdot 94\,\%\approx 91\,\%$ where $\eta\approx0.98$ is
the field overlap with
the ideal dipole mode\,\cite{golla2012}. But in our measurements,
we are not able to reach this limit. Our numerical simulations imply,
however, that we are currently not primarily limited by the covered
solid 
angle, but by the spatial extent of the ion and the optical
aberrations (see table \ref{tab:IonSize}). The optical aberrations
reduce the Strehl ratios for focusing from full solid angle and half
solid angle to a different degree, see Appendix for details. Since the
Strehl ratio is by far worse in the case of full solid angle
illumination, the half solid angle focusing yields the better coupling
efficiency. The measured coupling efficiency $G=13.7\pm 1.4\,\%$ is,
however, approximately twice as high as measured with our setup in
half solid angle configuration previously\,\cite{fischer_efficient_2014}. 
We conjecture that this is due to the fact that here we do not
fit a full saturation curve but restrict the experiment to low
saturation parameters, preventing an increase of the ion's spread in
position space and the associated stronger averaging over the
focal intensity distribution.
This effect was not considered in Ref.\,\cite{fischer_efficient_2014}.
Therefore, it is possible that the increase of the ion's extent at
higher excitation powers has affected the saturation of the ion there,
resulting in a larger saturation power and thus a seemingly smaller
coupling efficiency.
Furthermore, part of the improvement might be attributed to a
better preparation of the incident beam.

Enhancing the optical properties of our focusing system and
therefore also the coupling efficiency can be done by correcting for
the aberrations over the full aperture. The predominant aberrations in
our set-up are due to
form deviations of the mirror from the ideal parabolic shape. A higher
degree of form accuracy could be provided by including interferometric
measurement techniques\,\cite{leuchs_interferometric_2008} into the
mirror's production process. Alternatively, the aberrations can be
corrected by preshaping the incident wavefront before it enters the
parabolic mirror. Wavefront shaping techniques may rely on adaptive
optical elements (\textit{e.g.} liquid crystal display, deformable
mirror) or a (gray tone) phase plate. The latter technique has already
been tested for a 4$\pi$-PM of the same geometry as used
here\,\cite{golla2012}. Involving a second optical 
element for wavefront correction in front of the parabolic mirror
would only slightly add complexity to the system. If the corrective
element is reflection and refraction based, like a continuous membrane
deformable mirror is, the wavelength-independent character
of the imaging system is retained. 

The wavelength-independent character is the reason why the 4$\pi$-PM
is intrinsically free of chromatic aberrations. This is beneficial for
applications that require tight focusing not only for a monochromatic
light source. Typical applications can be found in the field of
microscopy: Confocal fluorescence microscopy\,\cite{sheppard_image_1977} usually requires correction for the
excitation and the detection wavelength; RESOLFT-type far-field
Nanoscopy\,\cite{hell_far-field_2007} in addition requires correction
for the depletion beam. Further examples are two- and three-photon
microscopy\,\cite{horton_vivo_2013, denk_two-photon_1990} or Raman
microscopy\,\cite{RamanMic}. The variety of reflective materials allows
to specialize the 4$\pi$-PM for a specific application, \textit{e.g.}
for high power applications, high sensitivity measurements or even for
applications where wavelengths from the deep UV to the far IR are used
at the same time. This ability may enable new illumination or imaging
techniques that are not possible with today's technology.

\section{Conclusion}

We experimentally characterized a light-matter interface based on a
4$\pi$-PM. By limiting its aperture, we could demonstrate an
effective excitation \textit{PSF} having a lateral spot size of $209\pm 20$\,nm
in vacuum. That corresponds to $0.57\cdot\lambda_{exc}$. Using the
full mirror we observed a strong splitting of the focal peak along the
axial direction 
due to form deviations of the parabolic mirror. Measuring the induced
aberrations interferometrically and including the results in numerical
simulations yields values that are consistent with the experiment. In
addition, we measured the light-matter coupling efficiency to be
$G=13.7 \pm 1.4\,\%$. This value is also in good agreement with our
simulations. Our findings allow us to infer that we are currently
limited by the aberrations of the parabolic mirror and the spatial
extent of the ion.   

We can surpass our current technical limitations by correcting the
aberrations of the 4$\pi$-PM. This would even further reduce the focal
spot size and increase the coupling efficiency. Ways for wavefront
correction were given in the discussion section of this work. We also
may apply higher trap frequencies or ground state cooling techniques
to reduce the spatial extent of the ion. Alternatively, we can trap
doubly ionized Ytterbium in our experimental set-up
\cite{heugel_resonant_2016}. When trapping a $^{174}\text{Yb}^{2+}$
ion we may not need to change the trapping or cooling techniques
because the ion's spatial extent is smaller in the Doppler limit
(higher charge, narrower transition linewidth). Furthermore,
$^{174}\text{Yb}^{2+}$ provides a closed two-level transition that is
desirable in many experiments on the fundamentals of light-matter
interaction. This may be a path for  
realizing a set-up capable of reaching the ultimate limitations of
focusing in free space\,\cite{lindlein_new_2007,sondermann_design_2007}.

\section*{Acknowledgements}
The authors thank S. Heugel, B. Srivathsan, I. Harder and M. Weber for
fruitful discussions and M. Weber for valuable comments on the
manuscript. 
G.L. acknowledges financial support from the European Research
Council via the Advanced Grant `PACART'.

\section{Appendix}

\subsection*{Simulating the expected coupling efficiency $G$}

We simulate the focal intensity distribution along the principal axes
of our system (x, y, z) based on a generalization of the method
presented in\,\cite{richards_electromagnetic_1959} including the
aberrations of our parabolic mirror.
The electric field, that the ion experiences is an
average\,\cite{teo2011} over the electric field in the focus by the spread of
the wavefunction of the ion due to its finite temperature. In order to
calculate the expected effective PSF of the PM we convolve the
simulated intensity distribution in the focus with the spatial extent
of the ion along the corresponding trap axis. In order to deduce the
coupling efficiency $G$, we compare the resulting intensity in the
focus to the one of a perfectly focused linear dipole wave
\cite{golla2012}, with equal ingoing power, in all three
directions. The product of these three ratios yields the expected
$G$. 

\subsection*{Influence of the spatial extent of the ion's wave function}

For determining the spatial extent of the Doppler cooled ion, we
assume it to be in a thermal state\,\cite{eschner_sub-wavelength_2003}.
Hence the spatial extent in each
dimension is described by a Gaussian shaped wave packet with width
$\sigma_{i},\,i\in (x,y,z)$. The width can by calculated from the
ground state wave packet $\sigma_{i,0}$ by
\cite{eschner_sub-wavelength_2003} 
\begin{equation}
\sigma_{i} = \sqrt{2\overline{n}_i+1}\, \sigma_{i,0}\, \text{,}
\nonumber
\end{equation} 
where $\overline{n}_i$ denotes the mean phonon number of the harmonic
oscillator in $i$-th direction and
$\sigma_{i,0}=\sqrt{\hbar/(2\,m\,\omega_i)}$. $m$ is the ion's mass
and $\omega_i$ the trap frequency in direction $i$. The probability
for finding the ion 
at $x_i$ is hence described by  
\begin{equation}
\left|\Psi_i (x_i)\right|^2 =
\frac{1}{\sigma_{i}\sqrt{2\pi}}e^{-\frac{1}{2}(\frac{x_i}{\sigma_{i}})^2}\text{.} 
\nonumber
\end{equation}
The mean phonon numbers are calculated using the semiclassical rate equations approach
\cite{stenholm1986,Eschner_LaserCooling}. For the common case of
cooling with laser beams impinging from small parts of the solid angle
the average number of motional quanta along trap axis $i$ can, in the
Lamb-Dicke regime, be approximated to be 
\begin{equation}
\overline{n}_i=\frac{\alpha
  \rho_{22}\left(\Delta,S\right)+\cos^2\vartheta
  \rho_{22}\left(\Delta-\omega_i,S\right)}{\cos^2\vartheta
  \left(\rho_{22}\left(\Delta+\omega_i,S\right)
  -\rho_{22}\left(\Delta-\omega_i,S\right)\right)},  
\label{eq:nbar}
\end{equation}
where
$\rho_{22}\left(\Delta,S\right)=S/2\left(1+\left(2\Delta/\Gamma\right)^2+S\right)$
is the upper level population with respect to detuning and saturation
parameter, $\vartheta$ the angle between the $i\textrm{-th}$ trap axis
with the $\vec{\textrm{k}}$-vector of the laser beam, and $\alpha$ a
factor depending on the emission pattern of the ion. For
$^{174}\textrm{Yb}^+$ ions the emission pattern is isotropic and
therefore $\alpha=1/3$\,\cite{stenholm1986}. In the case of cooling the
ion with a dipole wave, the cooling mode has a continuous spectrum of
$\vec{\textrm{k}}$-vectors, hence the factor determining the overlap
of the beam and the trap axis $\cos^2 \vartheta$ has to be averaged
over the incoming field linear dipole field, that is has the form
$\sin^2\vartheta$. With the geometry of the trap and focusing employed
in the setup described here, one of the trap axes is parallel to the
optical axis of the mirror, while the other two are perpendicular. The
mean overlap for these two cases, the averaged value of
\begin{align}
\eta_{\text{ax}} &= \iint d\vartheta\, d\varphi\,\frac{3}{8\pi}\sin^3
\vartheta \cos^2 \,(\vartheta + \pi/2) \nonumber\\ 
\eta_{\text{rad}} &= \iint d\vartheta\, d\varphi
\,\frac{3}{8\pi}\sin^3 \vartheta \cos^2 \,(\vartheta + 0) 
\label{eq:overlap}
\end{align}
with the integration along polar angle $\varphi$ and azimuthal angle
$\vartheta$. This leads to a mean overlap of $\eta_{\textrm{ax}}=1/5$
of the linear dipole mode with the axial trap axis and
$\eta_{\textrm{rad}}=2/5$ with the radial ones. 
For negligible excitation, a detuning of
$\Delta/2\pi=14.2\,\text{MHz}$ and the trap frequencies $\omega_x
/2\pi = 482.6\,\text{kHz}$, $\omega_y /2\pi = 491.7\,\text{kHz}$, and
$\omega_z /2\pi = 1025\,\text{kHz}$, respectively, this yields
$\overline{n}_{x,y} \approx 20 $, and $\overline{n}_z \approx 14$. The 
trap frequencies are determined by applying an AC signal to one of the
compensation electrodes and scanning the applied frequency over the
frequency range supposed to contain the trap frequencies while
monitoring the the rate of fluorescence photons. 

We only use excitation powers $P_{exc}$ such that $S\leq1/10$. This
equals a detection count rate of $R_{det}\leq 392\, cps$. For larger
excitation powers, we expect the spatial extent of the ion to be
comparable to the size of the focal intensity distribution and the
coupling efficiency to be reduced. Since $\overline{n}$ grows linearly
with $S$\,\cite{chang2014}, keeping $S\le0.1$ ensures that the spatial
extent of the ion is approximately constant over the whole measurement
range. A change of $10\,\%$ in $S$ corresponds to a change of
approximately $5\,\%$ in $G$.  

\subsection*{Interplay of mirror aberrations and focusing geometry}

\begin{figure}
\centering
\includegraphics[width=.95\linewidth]{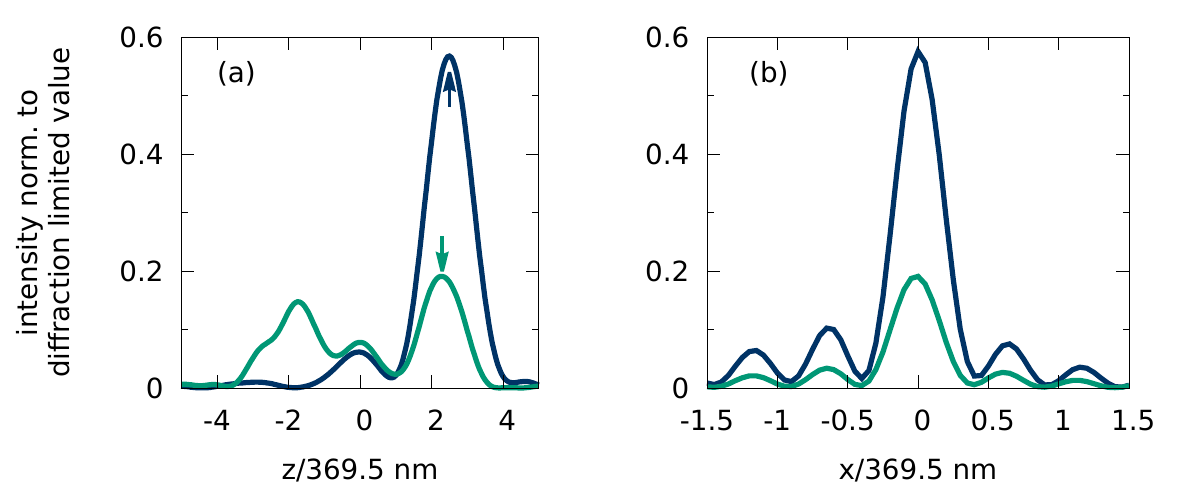}
 \caption{\label{fig:Strehl}
   Axial (a) and transverse (b) cuts through simulated focal
   intensity distributions. Green lines denote the case of focusing
   with the full mirror, i.e. covering 94\,\% of the solid angle
   relevant for a linear dipole. Blue lines represent the results for
   focusing from half solid angle as explained in the text. The
   transverse cuts in (b) are taken at the axial positions with
   maximum intensity as marked by the arrows in (a). The intensities
   are given relative to the ones obtained without any aberrations but
   at same solid angle. The spread of the ion's wave function is
   neglected in all simulations underlying this figure.} 
\end{figure}

To clarify the reasons for the counter-intuitive finding of a larger
coupling efficiency when focusing from only half solid angle, we
discuss the influence of the mirror's aberrations in more detail.
Here, focusing from half solid angle means that no light is incident
onto the parabolic mirror for radial distances to the optical axis
larger than twice the focal length of the parabola.
Figure\,\ref{fig:Strehl} shows the results of simulations of the focal
intensity distributions when only accounting for the aberrations of
the parabolic mirror but neglecting the trapped ion's wave function
finite extent.
For a parabolic mirror free of aberrations, i.e. a Strehl ratio of 1 no
matter which portion of the mirror is used for focusing, one would
expect coupling efficiencies of 90\% when using the full mirror and
48\% when focusing from half solid angle, assuming a mode overlap of
$\eta\approx0.98$ in both cases. 
But as is apparent from the data in Fig.\,\ref{fig:Strehl}, the Strehl
ratio drops by a factor of 3 when focusing with the full mirror
in comparison to using only half of the solid angle for focusing.
Therefore, the expected increase in coupling efficiency by focusing
with the full mirror is hindered by the large decrease in Strehl
ratio.
This particular, device-specific distribution of the aberrations
results in a larger coupling efficiency for the half-solid-angle case,
as it is found in our experiments.

\subsection*{Determination of the detection efficiency}

The main contributions to the detection efficiency $\eta_{det}$ are
the reflectivity of the parabolic mirror for the detection mode, the
quantum efficiency of the PMT, the covered solid angle of the
parabolic mirror, the reflectivity of an additional beam splitter, the
transmission of the dichroic mirror, and the transmission of the clean
up filters, respectively. The reflectivity of the parabolic mirror and
the quantum efficiency of the PMT are only known for a wavelength of
369.7\,nm and amount to 67\,\% and 13\,\%
\cite{maiwald_collecting_2012}, respectively. We expect this value to
be lower for the detection wavelength of 297.1\,nm. Together with the
covered solid angle of the parabolic mirror of 81\,\%
\cite{maiwald_collecting_2012} and the polarization averaged design
parameters for the beam splitter, dichroic mirror and the clean up
filters of 43\,\%, the detection efficiency has to be lower than
3\,\%. 

For a precise value, we additionally measure the detection efficiency
via pulsed excitation. We pump the ion from the ground state into the
metastable $\text{D}_{3/2}$ dark state by focusing a strong $30\,\mu$s
long laser pulse at a wavelength of $369.5$\, nm through the backside
hole of the parabolic mirror. After that, we drive the
$\text{D}_{3/2}\text{ - D[3/2]}_{1/2}$ transition with a strong laser
pulse for about $30\,\mu$s to ensure that the
$\text{D[3/2]}_{1/2}\text{ - S}_{1/2}$ decay takes place. During this
decay, a photon at the detection wavelength of $297.1$\,nm is
emitted. While applying the infrared light, no UV light is driving the
ion and only one detection photon can be emitted. For repeating the
experiment with a pulse sequence rate of 10\,kHz the background
corrected count rate amounts to 142\,cps. This yields the detection
efficiency $\eta_{det}=142/10000 = 1.4\,\%$.

%

\end{document}